\documentclass[aps,prl,preprint,showkeys]{revtex4}

\begin{document}

\font\brm=cmr12 scaled\magstep1

\title{\bf Simple observations concerning black holes and probability
\footnote{Essay awarded ``Honorable Mention'' in the Gravity 
Research Foundation 2009 Essay Competition}}

\author{S\'andor Hegyi\thinspace}

\email[E-mail: ]{hegyi@rmki.kfki.hu}

\affiliation{KFKI Research Institute for Particle and Nuclear Physics\\
             H-1525 Budapest, P.O. Box 49. Hungary}

\date{\today}
 
\begin{abstract}
What is common in a black hole and a bell-shaped curve? The question
does not seem to make much sense but it is argued that black holes and
the limit distributions of probability theory share several properties
when their entropy and information content are compared. In particular
the no-hair theorem, the entropy maximization and holographic bound,
and the quantization of entropy of black holes have their respective
analogues for stable limit distributions. This observation suggests
that the central limit theorem can play a fundamental role in black
hole statistical mechanics and in a possibly emergent nature of
gravity.
\end{abstract}


\maketitle

Ever since the pioneering work of Bekenstein [1] and Hawking [2] black
hole (BH) thermodynamics remained one of the most intensively studied
subjects in gravity research [3,4]. This can be attributed to the fact
that the thermal properties of BHs seem to have a fundamentally quantum
gravitational origin: quantities such as the temperature and entropy of
a black hole depend both on Newton's constant~$G$ and Planck's 
constant~$\hbar$.
Interestingly, the several different approaches to the quantum theory
of gravitation exhibit a profound universality [4]: at the microscopic
level these considerations attribute the BH entropy to different
microstates, nevertheless all of them correctly reproduce the
Bekenstein-Hawking formula $S_{\rm BH}=A/4$ relating the entropy 
$S_{\rm BH}$ of a black hole to the surface area~$A$ of its event 
horizon measured in Planck units.

A possible cause of this surprising and not fully explained success may
lie in the central limit theorem (CLT) of probability theory. The CLT
provides the core of the mathematical foundations of statistical
mechanics [5] and hence its utility in a
deeper understanding of BH entropy would not be totally unexpected.

The essence of CLT is not that foreign to the realm of black holes. BHs
are formed by the aggregation of masses until the limit of a
gravitational collapse is reached. The CLT deals with aggregation of
independent and identically distributed random variables and specifies
the probability distribution of the properly normalized aggregate in
the limit of an infinite number of its individual components [6]. The
CLT tells that the common distribution of the aggregated degrees of
freedom displays an overwhelming simplicity and universality.
Approaching the limit, the details of the individual
components' distribution progressively disappear and
finally one arrives at a probability law characterized only by a few
parameters. Moreover, the same limit law will emerge for infinitely
many different choices of the aggregated components. If the first two
moments of the individual distributions exist the limit law is the
Gaussian specified fully by its mean and variance.

The fact that the limit laws obey a very simple shape irrespective of
how complex is the distribution of the individual components is
reminiscent of the no-hair theorem: externally a black hole is
characterized only by its mass, electric charge and angular momentum.
Any other information about the collapsing matter such as quantum
numbers otherwise conserved in laboratory circumstances are not
preserved by BHs. In case of the CLT the higher order fluctuations of
the aggregated random variables are not preserved. These fluctuations,
however large but finite, are progressively smoothed out as the number
of individual components increases. As a result, for the emerging
Gaussian only the first two cumulant moments exist (mean and variance)
all the higher order cumulants characterizing higher fluctuations
vanish.

 The aggregation of random variables means convolution of their
distributions. Convolution is an information burning operation, the
entropy can not decrease under its repeated application. Therefore the
limit distributions must have larger entropy than the individual
components being aggregated. Indeed, the Gaussian law has maximum
entropy among the probability densities of fixed variance. This was
shown by Shannon in his classic paper setting the foundations of
information theory [7].

 From the no-hair theorem we know that there is an enormous
information loss when black holes are formed. Accordingly BHs have much
larger entropy than any known object of same mass. The entropy of black
holes scales with the surface area, $S\sim A$, whereas non-BH
objects satisfy the 't Hooft entropy bound $S<A^{3/4}$~[8].
Recently it was pointed out that in the classical limit the maximum
entropy of a quantized surface is proportional to its area for a wide
class of surface quantization schemes [9]. Thus the BH horizon has
maximum entropy among the spherical surfaces of fixed area. The
similarity to limit laws is obvious with their variance or
\textit{width} playing the role of BH event horizon \textit{area}. But
it is early to conclude that the analogy has solid foundations. The
possible existence of hyperentropic objects and so-called monster
configurations having larger entropy than black holes is still 
debated [10].

In close correspondence with entropy maximization by black holes various
entropy bounds were formulated for ordinary physical systems. The most
appropriate for our purposes is the holographic bound and its
derivation via the Susskind process [8,11]. Consider a spherical system
of entropy $S$ and surface area $A$ that includes gravity. Let us allow 
the system to collapse into a black hole. 
During the collapse the initial entropy $S$ increases to $S_{\rm BH}$
because of the second law of thermodynamics whereas the confining area
$A$ of the system decreases to the black hole event
horizon area $A_{\rm BH}$. That is, $S<S_{\rm BH}=A_{\rm BH}/4<A/4$
and therefore the entropy of any ordinary system is
ultimately bounded by its surface area, $S<A/4$.

The logic of the Susskind process can be applied to probability
distributions with the role of black holes replaced by limit laws.
Consider a random variable $x$ with probability density $f(x)$
of finite width but the variance should not necessarily
exist. We are interested in the upper bound on the Shannon
entropy $S(x)=-\int f(x)\,\ln f(x)\,dx$.
Making an infinite convolution of $f(x)$ with
itself while the width being kept fixed a Gaussian or L\'evy limit law
emerges. Its Shannon entropy is certainly larger than that of
\textit{f(x)} since entropy can not decrease under convolution. A
unique property of limit laws is \textit{stability}: up to a
location and scale change their shape remains unmodified under
convolution. Since entropy is shift invariant, it is the scale
parameter, i.e., the width what represents the Shannon entropy of limit
distributions. Therefore $S(x)$ of an arbitrarily chosen
probability density is bounded by its width.

Due to the property $S(\lambda x)=S(x)+\log\lambda$
of Shannon entropy it is actually the logarithmic
width what measures the entropy of stable laws. The $N$-fold
convolution of a $\lambda=1$ stable distribution changes
its scale parameter to $\lambda=N^{1/\alpha}$ where $\alpha$ is the
index of stability, $0<\alpha\leq2$. The $\alpha=2$ case is the 
Gaussian, otherwise we have a L\'evy law with inverse power law tail
$x^{-1-\alpha}$ [6]. While the width of limit distributions measures the 
number of stable degrees of freedom being aggregated, the surface area 
of black holes counts the number of quantum mechanical degrees of 
freedom on the event horizon [8,11].

The idea that the black hole event horizon might be quantized was
suggested by Bekenstein already in the first years of research of BH
thermodynamics~[12]. In his intuitive picture the horizon surface is
tiled with plaquettes of area of the squared Planck length. Although
the details of the recent approaches to surface quantization are
different, each consistently specifies a quantum of area on the event
horizon and hence a quantum of entropy being proportional to it.

A somewhat analogous property characterizes the \textit{discrete} limit
distributions. Aggregating discrete random variables of finite mean the
emerging limit law is the Poissonian. The $N$-fold convolution
of a Poisson distribution of unit mean reproduces it with mean $N$.
Thus the mean of the Poisson law measures the number of
aggregated degrees of freedom. Importantly, this holds for
\textit{non}-Poissonian discrete limit laws as well. They can 
be decomposed into a Poisson distribution of identical clusters. Under
convolution \textit{only} the Poissonian component gets modified (its
mean increases) the probability distribution inside the clusters and
hence its Shannon entropy remains unchanged. This behavior is the
discrete analogue of stability [13]. Therefore the entropy of a
non{}-Poissonian limit law can be thought as quantized: its increase
under convolution is due to an increasing number of independent and
identical clusters of fixed Shannon entropy. A possible source of 
discrete stability is the contamination of L\'evy fluctuations by
Poisson noise.

The distribution $p(k)$ inside the clusters has a power law tail
$k^{-1-\alpha}$. The index $\alpha$ is now restricted to $0<\alpha<1$
and it yields the probability $p(k=1)$. In the limit $\alpha\to1$ the 
pure Poissonian is recovered as the limit distribution. The Shannon 
entropy of $p(k)$ is fixed by the index $\alpha$ being inversely
proportional to it. The role of $\alpha$ bears some resemblance 
with the Immirzi parameter determining the size of area quanta 
in the loop quantum gravity approach to black hole entropy [14].

We have seen that the stability property of limit distributions plays a
central role in establishing the correspondence with black holes. For
continuous limit laws stability expresses a scale symmetry of the same
kind what is encountered in renormalization group methods applied to
phase transitions [15]. Aggregating random variables is the equivalent
of forming block spins whereas scale invariance of the aggregate
represents fixed point behavior near the critical point.
Therefore the stability of limit distributions is intimately related
to the observed similarities between BHs and phase transitions [16,17]. 
Note also that the L\'evy limit laws (continuous or discrete) have 
power law tails, in accordance with self-similar fluctuations of 
the event horizon's shape which can be responsible of BH entropy~[18]. 
It is worth considering that two key concepts of 
the central limit theorem, the $N\to\infty$ limit of aggregated 
random variables and the stability of the aggregate's distribution 
are patterns parallel to the  $N\to\infty$ limit of $SU(N)$ gauge
theories and the holographic nature of BH entropy. The latter are 
two fundamental ideas behind the AdS/CFT correspondence. Thus one 
is led to speculate that a probabilistic view of the holographic 
principle may be also formulated on the basis of the CLT.

Recently it was suggested that a near-horizon conformal symmetry governs 
black hole thermodynamics, independently of the details of quantum 
gravity [4,19]. This picture reveals an emergent character of the 
thermal properties of BHs. The correspondence between limit laws and 
black holes may ultimately be rooted in an emergent nature of 
gravity~[17,20] with the central limit theorem playing a defining 
role in it. A less ambitious but more definite view of our result is 
that, similarly to Hawking radiation, black hole entropy also has 
analogues in physics outside general relativity, especially in 
fluctuation phenomena which give birth of stable probability 
distributions. In any case, it will be nice to see if a new and
fruitful interaction develops between probability theory
and gravity research.


\begin{thebibliography}{99}
\bibitem{[1]} J.D. Bekenstein, Phys. Rev. D 7, 2333 (1973)
\bibitem{[2]} S.W. Hawking, Commun. Math. Phys. 43, 199 (1975)
\bibitem{[3]} R.M. Wald, Living Rev. Rel. 4, 6 (2001)
\bibitem{[4]} S. Carlip, Lect. Notes Phys. 769, 89 (2009)
\bibitem{[5]} A.I. Khinchin, \textit{Mathematical Foundations of 
              Statistical Mechanics}, Dover, New York (1949)
\bibitem{[6]} B.V. Gnedenko and A.N. Kolmogorov, \textit{Limit 
              Distributions of Sums of Independent Random Variables},
              Addison-Wesley, Cambridge (1954)
\bibitem{[7]} C.E. Shannon, Bell Syst. Tech. J. 27, 349 (1948)
\bibitem{[8]} G. 't Hooft, in \textit{Salamfestschrift},
              ed. by A. Ali, J. Ellis and S. Randjbar-Daemi,
              World-Scientific, Singapore (1994) and 
              {\tt arXiv:gr-qc/9310026}
\bibitem{[9]} R.V. Korkin and I.B. Khriplovich, JETP \ 95, 1 (2002)
\bibitem{[10]}  
  D. Marolf and R.D. Sorkin, Phys. Rev. D 69, 024014 (2004) \\
  J.D. Bekenstein, Phys. Rev. D 70, 121502 (2004) \\
  S.D.H. Hsu and D. Reeb, Phys. Lett. B 658, 244 (2008) \\
  R.D. Sorkin, R.M. Wald and Z.J. Zhang, Gen. Rel. Grav. 13, 1127 (1981)
\bibitem{[11]} L. Susskind, J. Math. Phys. 36, 6377 (1995)
\bibitem{[12]} J.D. Bekenstein, Lett. Nuovo Cim. 11, 467 (1974)
\bibitem{[13]} F.W. Steutel and K. Van Harn, Ann. Probab. 7, 893 (1979)
\bibitem{[14]} A. Ashtekar, J. Baez, A. Corichi and K. Krasnov, Phys. 
  Rev. Lett. 80, 904~(1998)
\bibitem{[15]} G. Jona{}-Lasinio, Nuovo Cim. B 26, 99 (1975)
\bibitem{[16]} L. Susskind, L. Thorlacius and J. Uglum,
               Phys. Rev. D 48, 3743 (1993)
\bibitem{[17]} R.B. Laughlin, Int. J. Mod. Phys. A 18, 831 (2003)
\bibitem{[18]} R.D. Sorkin, in \textit{Proc. First Australasian
               Conference on General Relativity and Gravitation}, 
               ed. by D. Wiltshire, University of Adelaide Press, 
               Adelaide (1996) and {\tt arXiv:gr-qc/9701056}
\bibitem{[19]} S. Carlip, Gen. Rel. Grav. 39, 1519 (2007)
\bibitem{[20]} O. Dreyer, PoSQG-Ph:016,2007 
               and {\tt arXiv:0710.4350}
\end{thebibliography}
\end{document}